\documentstyle[multicol,prl,aps,epsf]{revtex}
\begin{document}

\begin{multicols}{2}
\noindent{
\large\bf{Reply to ``Comment on Evidence for the droplet picture of
spin glasses''}}
\smallskip
\narrowtext
\parskip 0cm Using Monte Carlo simulations (MCS) and the
Migdal-Kadanoff approximation (MKA), Marinari et al. study in their
comment \cite{marinari} on our paper\cite{moore} the link overlap
between two replicas of a three--dimensional Ising spin glass in the
presence of a coupling between the replicas.  They claim that the
results of the MCS indicate replica symmetry breaking (RSB), while
those of the MKA are trivial, and that moderate size lattices display
the true low temperature behavior.  Here we show that these claims are
incorrect, and that the results of MCS and MKA both can be explained
within the droplet picture.

The link overlap is defined as $q^{(L)}(\epsilon)=(1/3V)\sum\langle
\sigma_i \sigma_{j} \tau_i \tau_{j}\rangle$ where the sum is over all
nearest-neighbor pairs $\{ij\}$, and the brackets denote the thermal
and disorder average. $\sigma$ and $\tau$ denote the spins in the two
replicas. The Hamiltonian used for the evaluation of the thermodynamic
average is $H[\sigma,\tau] = H_0 [\sigma]+H_0 [\tau]- \epsilon\sum
\sigma_i \sigma_{j} \tau_i \tau_{j}$, where $H_0$ is the ordinary spin
glass Hamiltonian. For the subsequent discussion, it is useful to write
$q^{(\infty)}(\epsilon)$ in the form $q^{(\infty)}(\epsilon)=q_{+} + A_{+}
|\epsilon|^{\lambda_+}$ for $\epsilon > 0$ and $q^{(\infty)}(\epsilon)=q_{-} +
A_{-}|\epsilon|^{\lambda_-}$ for $\epsilon < 0$. 

In the mean-field RSB picture, $q_{+}> q_{-}$, and
$\lambda_+=\lambda_-=1/2$, and Marinari et al claim to see a trend towards this
discontinuous behavior in their MCS data (Fig.1 of \cite{marinari}).
Alternatively, if they assume continuous behavior, they find a
 value ${\lambda_\pm} \simeq 0.25$.  These conclusions are
based on the assumptions that there are no corrections to the pure power-law behavior, and that $\lambda_+=\lambda_-$.  However,
neither assumption is justified, and the most natural interpretation of Fig.1 of \cite{marinari} is $q_{+}= q_{-}$, and ${\lambda_\pm} \simeq 1/2$.

This result, as well as the results of the MKA, is in fact fully
compatible with the droplet picture. Using scaling arguments similar
to those in \cite{bray}, the value of $\lambda_-$ and $\lambda_+$ at
low temperatures can be derived in the following way: The energy cost
of the formation of a spin-flipped ``droplet'' of radius $l$ in one of
the replicas is of the order $l^\theta+\epsilon l^{d_s}$, where $d_s$
denotes the fractal dimension of the droplet surface, and $\theta$ is
the scaling dimension for the domain wall energy.  For negative
$\epsilon$, droplets of a characteristic size $ l^* \sim
(1/|\epsilon|)^{1/(d_s-\theta)}$ are formed, since they lower the
energy of the system. Since flipping a cluster only affects links on
the surface of the cluster, this gives $q^{(\infty)}(\epsilon)\simeq
q^{(\infty)}(0) - C |\epsilon|^{(d-d_s)/(d_s-\theta)}$.  Within the MKA
$d_s=d-1$ and $\theta \simeq 0.24$, leading to $\lambda_- \simeq
0.57$. For a cubic lattice, one has $\theta \simeq 0.2$, and $d_s
\simeq 2.2$, leading to $\lambda_-\simeq 0.4$. For positive
$\epsilon$, the leading correction to the link overlap comes from the
suppression of the thermal excitation of large droplets and has for
low temperatures the form $q^{(\infty)}(\epsilon)-q^{(\infty)}(0)
\simeq k_BT(\epsilon/k_BT)^{(d+\theta-d_s)/d_s}$, leading to
$\lambda_+=(d+\theta-d_s)/d_s$. Its value in MKA is $\lambda_+ \simeq
0.62$, very close to $\lambda_-$.

 For finite temperatures and small systems, there are corrections to
this asymptotic behavior due to finite-size effects which replace the
nonanalyticity at $\epsilon=0$ with a linear behavior for small
$|\epsilon|$, and due to the influence of the critical fixed point,
where the leading behavior is linear in $\epsilon$. As we have argued
in \cite{moore}, the influence of the critical fixed point changes the
apparent value of the low-temperature exponents for the system sizes
studied in the MCS and the MKA.  The MCS data shown in \cite{marinari}
with an apparent value of 0.5 for ${\lambda_\pm}$ are fully compatible
with these predictions of the droplet picture. For the MKA, the
apparent exponent at $0.7T_c$ is close to 1 for $L\simeq 16$, leading
to the ``trivial'' behavior found in \cite{marinari}. However, at
lower temperatures, for the same small system sizes the
above-mentioned nontrivial features predicted by the droplet picture
become clearly visible, as shown in Fig.1.

\begin{figure}
\epsfysize=0.5\columnwidth{{\epsfbox{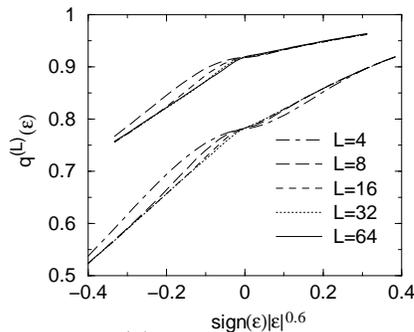}}}
\narrowtext{\caption{$q^{(L)}(\epsilon)$ in MKA at $T\simeq 0.38 T_c$ (bottom) and $0.14T_c$ (top) as
function of sign$(\epsilon)$$|\epsilon|^{0.6}$ for various system
sizes.}}
\end{figure}    

The apparent system size dependence of the exponents $\lambda_\pm $ allows us even to
estimate numerically the system sizes needed to see the true low
temperature scaling behavior. By iterating the recursion relations for
the coupling constants within MKA, we find that these system sizes are
of the order $L\simeq 1000$ at $T \simeq 0.7T_c$. We expect that similar system sizes would be needed for MCS to see the scaling behavior predicted by the droplet picture. 

This work was supported by EPSRC Grants GR/K79307 and GR/L38578.

\medskip

\noindent
H. Bokil,  A.J. Bray, B. Drossel, M.A. Moore

{\small

Department of Physics and Astronomy

University of Manchester

Manchester M13 9PL, U.K.

}

 \end{multicols} \end{document}